\documentclass[runningheads]{llncs}
\usepackage{graphicx}
\usepackage{xcolor}
\usepackage{amsmath}
\usepackage{algorithm}
\usepackage{algpseudocode}
\usepackage{amssymb}
\usepackage{multirow}
\usepackage{subfig}
\usepackage{makecell}
\usepackage{caption}
\usepackage{enumitem}
\usepackage{subfig}
\usepackage{textcomp}
\usepackage{hyperref}

\begin{document}
\title{P3GNN: A Privacy-Preserving Provenance Graph-Based Model for APT Detection in Software Defined Networking}
%
%

\author{Hedyeh Nazari\inst{1}\and
Abbas Yazdinejad\inst{1}\and
Ali Dehghantanha\inst{1}\and
Fattane Zarrinkalam\inst{2}\and
Gautam Srivastava\inst{3,4}}

\authorrunning{H. Nazari et al.}
%
\institute{Cyber Science Lab, Canada Cyber Foundry, University of Guelph, Canada\\
\email{hnazari@uoguelph.ca, ayazdine@uoguelph.ca, adehghan@uoguelph.ca} \and
Dept. of Engineering, University of Guelph, Canada\\
\email{fzarrink@uoguelph.ca}
\and
Dept. of Math and Computer Science, Brandon University, Manitoba, Canada\\
\email{srivastavag@brandonu.ca} \and
Research Centre for Interneural Computing, China Medical University, Taichung, Taiwan\\
\email{srivastavag@brandonu.ca}}

\maketitle              
\let\thefootnote\relax\footnotetext{This paper has been submitted to the ACM CCS 2024 Conference, AutonomousCyber 2024 Workshop for peer review and potential publication.}
\begin{abstract}

Software Defined Networking (SDN) has brought significant advancements in network management and programmability. However, this evolution has also heightened vulnerability to Advanced Persistent Threats (APTs), sophisticated and stealthy cyberattacks that traditional detection methods often fail to counter, especially in the face of zero-day exploits. A prevalent issue is the inadequacy of existing strategies to detect novel threats while addressing data privacy concerns in collaborative learning scenarios. This paper presents P3GNN (privacy-preserving provenance graph-based graph neural network (GNN) model), a novel model that synergizes Federated Learning (FL) with Graph Convolutional Networks (GCN) for effective APT detection in SDN environments. P3GNN utilizes unsupervised learning to analyze operational patterns within provenance graphs, identifying deviations indicative of security breaches. Its core feature is the integration of FL with homomorphic encryption, which fortifies data confidentiality and gradient integrity during collaborative learning. This approach addresses the critical challenge of data privacy in shared learning contexts. Key innovations of P3GNN include its ability to detect anomalies at the node level within provenance graphs, offering a detailed view of attack trajectories and enhancing security analysis. Furthermore, the model's unsupervised learning capability enables it to identify zero-day attacks by learning standard operational patterns. Empirical evaluation using the DARPA TCE3 dataset demonstrates P3GNN’s exceptional performance, achieving an accuracy of 0.93 and a low false positive rate of 0.06. 

\keywords{Advanced Persistent Threat, SDN, Graph Neural Network, Federated Learning.}
\end{abstract}

\section{Introduction}\label{Introduction}
SDN represents a novel approach in network architecture that aims to address prevailing challenges by separating the control plane from the data plane and using protocols such as OpenFlow \cite{6587999,cv1} for communication between these planes \cite{s1}. This approach turns network switches into simple packet handlers, with decision-making centralized in a software controller. The controller offers a comprehensive overview of the network and programming abstractions, enabling a centralized management system. This system grants operators programmatic and real-time control over networks and devices, simplifying network management and introducing flexibility that was previously lacking \cite{LATIF2020102563,7120048}.  However, the adoption of SDN introduces new security concerns, such as APT \cite{10.1145/3199478.3199481,cv3}. APTs pose an immense and formidable threat to cyberspace security. Characterized by their stealth and slow infiltration, APTs are typically launched by well-resourced groups, like state actors, and employ advanced tactics, techniques, and procedures (TTPs) for long-term network presence, targeting espionage or sabotage \cite{8606252}. As APTs become more sophisticated, cybersecurity solutions must become more dynamic and adaptable. 


Diverse strategies utilizing log files to counter APTs have evolved due to the limitations of traditional signature-based detection systems, particularly against zero-day exploits and advanced polymorphic malware \cite{4674371}. Among these, Host-based Intrusion Detection Systems (HIDS) stand out for their efficacy. Further, integrating HIDS into SDN networks appears promising for enhanced security \cite{YAZDINEJADNA2021107688}. Building on HIDS, the Provenance-based Intrusion Detection System (PIDS), especially its provenance graph method, utilizes provenance data's contextual information to pinpoint APT attacks \cite{wu2022paradise}. In single-host environments, a system provenance graph, structured as a directed acyclic graph (DAG), categorizes log file elements like processes, files, and sockets as vertices and their interactions as edges. This method effectively maps system processes, illustrating control and data flow between entities\cite{9900181,pasquier2017practical}. Previous research on log provenance graphs for APT detection faces limitations, particularly with misuse-based methods' inability to detect zero-day attacks. These methods, which compare provenance graphs to known attack graphs, struggle to identify new threats not captured in existing cyber threat reports \cite{10.1145/3319535.3363217}. In response, unsupervised anomaly-based methods offer a promising alternative by focusing on deviations from normal behavior within provenance graphs, essential for detecting new and emerging threats, including zero-day attacks \cite{s3}.

Previous research on detecting APTs using provenance graphs highlights significant challenges due to the complexity of these dense networks that represent system interactions. GNNs, in contrast to standard Deep Neural Networks (DNNs), are better suited for these tasks due to their ability to interpret graph-structured data, enabling them to not only detect but also visualize malicious activities within network contexts \cite{manzoor2016fast,LIANG2021168}. However, the effectiveness of GNNs is often limited by the availability of extensive log datasets necessary for training. Despite the potential of centralized training to mitigate data scarcity by pooling resources from multiple sources, practical implementation is hindered by privacy and security concerns, preventing organizations from sharing sensitive log data \cite{DBLP:journals/corr/GhiasvandC17,p2}.

Considering these challenges, in this paper, we introduce P3GNN, an explainable privacy-preserving model specifically designed for APT detection. P3GNN innovatively combines the principles of FL with GCN, using unsupervised learning to examine operational patterns in provenance graphs for identifying deviations that signal security breaches \cite{p3}. A key aspect of P3GNN is its strong focus on privacy and security. By integrating FL with homomorphic encryption, specifically the Paillier homomorphic cryptosystem (PHC), it ensures the confidentiality and integrity of data and gradients during collaborative learning, offering dual-layer protection for sensitive information \cite{konečný2017federated}. This integration is vital because, while FL allows for collaborative model training across different organizations without direct data sharing, it inherently provides limited privacy protection and doesn't fully resolve security issues, especially regarding data shared with third-party servers \cite{shokri2015privacy}. The concern of an untrustworthy server exploiting uploaded parameters to infer feature values from private training datasets is effectively addressed by the addition of partially homomorphic encryption in P3GNN. Additionally, P3GNN's capability to detect APTs at the node level within provenance graphs, identify anomalous nodes, and trace the full attack trajectory, is crucial for security analysts. This functionality offers a comprehensive view of attack paths and enhances the analysis of APT incidents. The main contributions of this paper can be summarized as follows:
\begin{itemize}
    \item {We propose a novel interpretable, privacy-preserving provenance graph-based FL model P3GNN for detecting APTs. Our model leverages GCN to adeptly navigate the intricate interconnections present in log provenance graphs. }
    \item {Our model operates in an unsupervised manner, effectively learning the standard operational patterns within provenance graphs and thereby identifying any deviations from the system's typical behavior. This capability enables our system to detect anomalies that could signify zero-day attacks, which are particularly challenging to identify due to their novel and previously unrecorded nature. }
    \item {The integration of FL with homomorphic encryption in P3GNN maintains the confidentiality of data and gradients in collaborative learning, addressing data sharing challenges and protecting log owners' privacy. This allows training on diverse datasets, including attack samples from multiple sources, without transferring sensitive log data. Training is conducted locally, leveraging distributed data while ensuring privacy and security.}
    \item {P3GNN can pinpoint anomalous nodes within provenance graphs at the node level. This feature significantly enhances the model's capability to trace the entire trajectory of APT attacks, providing valuable insights for security analysis. This attribute of P3GNN addresses a common issue in the field of artificial intelligence regarding interpretability.}
\end{itemize}
The following is the specification of the threat model we consider in this paper. 
\begin{itemize}
    \item \textbf{High Volume of Log Data with Subtle Anomalies:}
    The main challenge in detecting APTs within SDNs lies in the sheer volume of log data, where malicious activity can easily masquerade as normal system behavior. This makes it challenging to identify threats without examining the interconnections among various system events. Furthermore, the subtlety of such activities necessitates deep analysis. Advanced ML techniques are crucial for effectively identifying these hidden threats.
    \item \textbf{Privacy Risk in Data Sharing for GNN Training:} In a decentralized infrastructure for GNN training, sharing sensitive log data with external entities poses a significant privacy risk, potentially exposing critical operational details. 
    \item \textbf{Confidentiality of Gradients in FL setup:} The confidentiality of gradients is a concern, as an honest but curious server could potentially access sensitive user information through shared gradients and model parameters. An adversary can gain access to the private logs of an organization through gradient inversion attacks \cite{liang2023egia}.
    \end{itemize}
    By achieving our contributions, we address the following design goals in this research: 
\begin{itemize}
\item \textbf{Enhanced Anomaly Detection in Large Log Data:} To tackle the challenge of detecting APTs within vast log datasets, our solution employs an anomaly-based detection system using provenance graphs. These graphs help interpret the relationships and context within system logs, identifying deviations from normal behavior. We use GNNs to effectively handle and analyze these complexities, allowing for detailed tracking of anomalies at the node level and enabling quick investigations.

\item \textbf{Privacy-Preserving Data Handling for GNN Training:} In response to the privacy risks associated with data sharing in a decentralized GNN training infrastructure, our model leverages FL. This approach allows the training data to remain within the confines of the organization’s secure environment, thereby negating the need to transfer sensitive information externally. This method ensures that privacy is maintained during the GNN training process.

\item \textbf{Secure Gradient Sharing:} To mitigate the risk of gradient inversion attacks and protect user data, the model incorporates PHC for encrypting local gradients before transmission to the server. This encryption step is crucial in maintaining the confidentiality of sensitive information throughout the gradient-sharing phase, ensuring robust protection against potential adversarial access.
\end{itemize}
The rest of this paper is organized as follows. Section \ref{sec:Related Work} reviews the related literature. Section \ref{sec:proposed model} introduces the structure of the P3GNN model. Section \ref{sec:Performance Evaluation} evaluates the proposed model, compares the results to related works, and provides interpretability and complexity analysis for the model. Lastly, we conclude the paper in section \ref{sec:conclusion}.

\section{Related Work}\label{sec:Related Work}
APT detection has been the subject of numerous studies. In provenance-based methods, Milajerdi et al. \cite{8835390} developed Holmes, a real-time APT detection system that utilizes provenance graphs for correlating suspicious information flows. It employs expert knowledge of TTPs to match defined exploits within these graphs. This integration allows Holmes to effectively summarize attackers' actions through robust detection signals and high-level graphs, offering precise detection and low false alarm rates in real-world APT scenarios. Also, in \cite{8450016}, Xie et al. proposed a model named Pangoda, designed for large data environments. It analyzes provenance graphs to detect anomalies, balancing the focus between individual paths and the overall graph structure. This enables quick identification of intrusions and improves detection accuracy. In \cite{wang2020you}, the authors introduce ProvDetector, a malware detection tool that utilizes provenance graphs. ProvDetector operates by embedding paths within a provenance graph and applying the Local Outlier Factor method to identify malware effectively. Poirot, as proposed in \cite{10.1145/3319535.3363217}, is another provenance-based system designed for threat hunting. It focuses on correlating indicators identified by other systems and constructs attack graphs using expert knowledge from existing cyber threat reports. The system matches the provenance graphs with these attack graphs to detect threats, but it faces challenges in identifying unknown threats not already included in the TTPs and cyber threat reports. DeepTaskAPT introduced in \cite{9724394} is an LSTM-based deep learning model for detecting APTs by analyzing event sequences in the provenance graph. This model identifies unusual activities by being trained solely on benign sequences, which are created using a tree-based method for task generation. In \cite{han2020unicorn}, a system named UNICORN is presented for detecting APT anomalies using sketch-based, time-weighted provenance encoding and graph analysis. It focuses on supply-chain attacks, with multi-hop exploration and evolutionary modeling enhancing accuracy and efficiency.

In the context of APT detection in SDNs, the authors in \cite{shan2017apt} introduce a Hidden Markov Model-based APT detection method that effectively detects and analyzes multi-stage APTs in SDN, offering accurate threat identification with minimal overhead. In \cite{10.1145/3199478.3199481}, they propose a method that employs an attack tree-based model for SDN, enabling the detection of APTs by correlating multiple attack behaviors to establish an attack path. However, these methods fail to take into account the privacy of log owners. Huynh Thai Thi et al. present an FL approach for enhancing cyber threat detection in SDN. This method leverages collaborative intelligence across multiple parties while preserving data privacy. The technique focuses on proactive APT attack detection, utilizing FL to improve the performance of detection systems against unknown threats. For the IDS model, the authors employed three local models, including Gated Recurrent Unit (GRU), Long Short-Term Memory (LSTM), and Convolutional Neural Network (CNN) \cite{9931222}. In \cite{thi2023xfedhunter}, the XFedHunter model uses the FedAvg algorithm for decentralized APT detection in SDNs, enhancing privacy. It integrates CNN, GRU, and E-GraphSAGE for intrusion detection and employs the SHAP framework for model explainability. While this approach focuses on privacy by design, it does not explicitly protect model parameters like gradients and weights during training.


\section{Proposed Model} \label{sec:proposed model}
\begin{figure}
    \centering
    \includegraphics[scale=0.38]{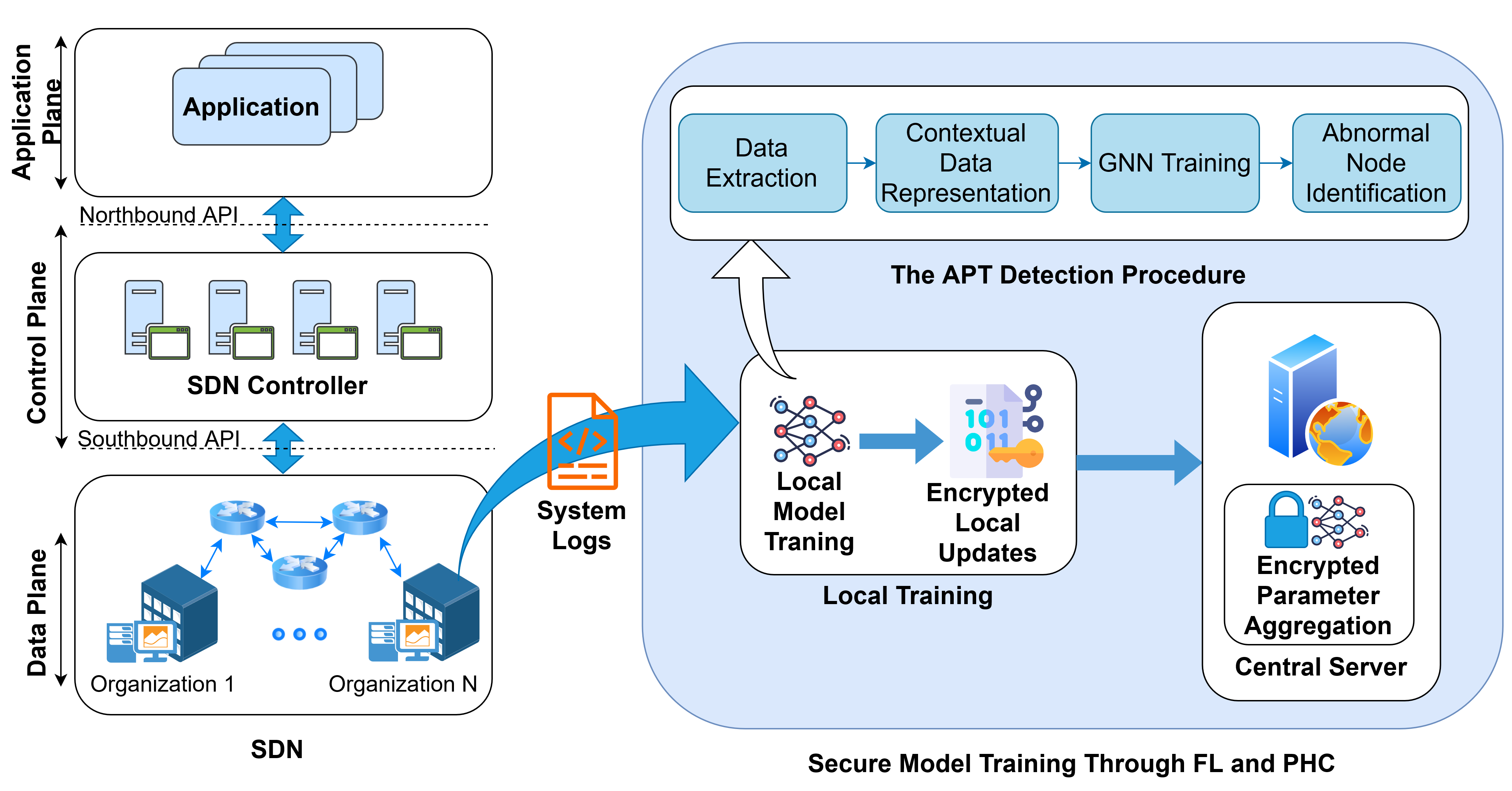}
    \caption{The architecture of P3GNN}
    \label{fig:architecture}
\end{figure}

In this section, we present an in-depth view of our model, P3GNN, specifically designed for detecting APTs in SDNs. The P3GNN model integrates three key technological paradigms: the APT detection procedure, the SDN infrastructure, and, the collaborative and privacy-preserving aspects of FL. Each of these components plays a pivotal role in creating a robust, efficient, and privacy-preserving APT detection system. Figure \ref{fig:architecture} shows the architecture of the different components in P3GNN. The following subsections will explain each component's contributions as well as how they work together to enhance our model's effectiveness and security.

\subsection{SDN Infrastructure}
In the design of our system, we view the SDN network as a potential zone for detecting malicious activities, which are likely to appear within the network's flow data. 
SDN encompasses three planes: the Control Plane, responsible for traffic direction decisions, often on a network operating system; the Data Plane, comprising physical or virtual routers and switches that forward packets per control plane instructions; and the management plane, overseeing network administration, monitoring, and policy enforcement. SDN utilizes northbound APIs for communication between the control plane and higher-level applications, enabling network programming, and southbound APIs (like OpenFlow) for connecting the control plane to data plane devices for management. To effectively monitor and scrutinize this data for any signs of such activities, it's essential to first set up the SDN network switches. A fundamental component of this process involves the data plane functionality of SDN switching devices, which are pivotal in packet forwarding. The data plane in an SDN switch operates distinctly compared to traditional networking devices. While legacy networks primarily rely on forwarding packets based on IP or MAC addresses, SDN switches introduce a more versatile approach \cite{xia2014survey}. These switches are tasked with forwarding network flow data for APT detection analysis via the OpenFlow protocol.

\subsection{APT Detection Procedure in P3GNN}
\begin{figure}
    \centering
    \includegraphics[scale=0.23]{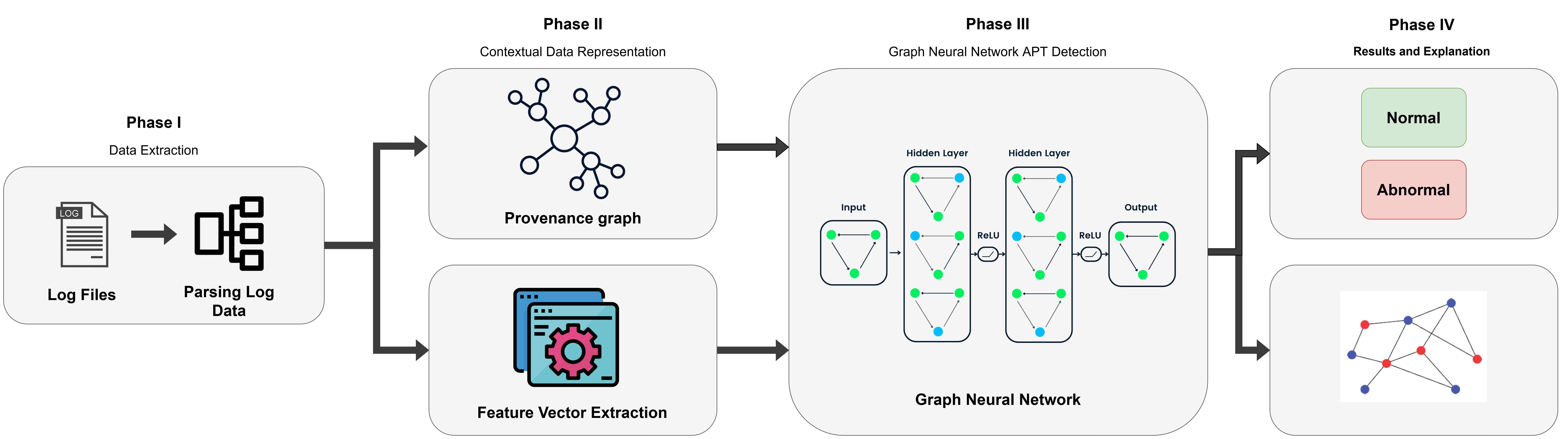}
    \caption{APT detection procedure}
    \label{fig:phases}
\end{figure}

The APT detection procedure of the P3GNN model as shown in Figure \ref{fig:phases} is structured into four phases: data extraction, contextual data representation, GNN training, and finally abnormal node identification and analysis. The subsequent sections will provide an in-depth exploration of each phase, detailing their functionalities and contributions to the model's capability in APT detection.

\subsubsection{Data extraction:}

In the initial phase of our model, we focus on the extraction and initial preprocessing of log data. The primary source of our data comprises system logs, encompassing a wide range of information such as timestamps, process names, and execution paths. Utilizing regular expression techniques, we parse these logs to transform the raw data into a structured format. We define \( \Sigma \) as the set of all unique symbols present in the log data. Using regular expressions, we create definitions for specific patterns found in log entries. For example, to identify a UUID, we use a regular expression denoted as \( R_{\text{uuid}} \). This expression is defined as \( \text{"uuid\":\""} \cdot (\Sigma^*) \cdot \text{"\""} \), where \( \Sigma^* \) represents any sequence of characters from our defined symbol set.

The process of transforming log entries into structured data is encapsulated by the parsing function \( \text{ParseLog} \). Let \( L \) represent the collection of all log entries. Each entry \( l \) in \( L \) is a tuple containing elements like timestamp, process name, and execution path. The function \( \text{ParseLog} \) applies the defined regular expressions to each log entry \( l \) from \( L \), converting it into a structured data format represented by \( d \) in the set \( D \). This transformation can be formally expressed as:
\begin{equation}
    \text{ParseLog}(l) : L \rightarrow D 
\end{equation}

This methodical approach allows us to efficiently convert unstructured log data into a format that is more suitable for analysis and further processing.
\subsubsection{Contextual data representation:}
Our approach to detecting anomalous network nodes thrives in environments without predefined attack patterns, moving away from traditional supervised learning that relies on clear labels, which are often unavailable due to the lack of specific attack knowledge during training. We utilize a GCN designed to differentiate and comprehend the subtle behaviors of benign nodes. The core idea is that nodes involved in malicious activities exhibit distinct behavioral anomalies compared to benign ones, allowing our model to learn and identify unique patterns of relationships among node types. The feature extraction process involves an approach wherein each node's interactions and connections are translated into feature vectors. Thus, the contextual data representation phase is as follows.
Our primary objective is to construct a provenance graph and extract features from the preprocessed logs. The provenance graph, denoted as \( G = (V, E) \), is built based on the relationships identified in the logs. Here, \( V \) represents the set of vertices (entities such as files, processes, sockets), and \( E \) represents the set of edges (interactions like read, write, fork).

We define $N_v$ as the total number of unique node types and $N_e$  as the total number of unique edge types present in $G$. To facilitate the mapping of node and edge types to unique integers, we introduce two mapping functions: \( M_v: \text{Types} \rightarrow \{0, \ldots, N_v-1\} \) for nodes, and \( M_e: \text{Types} \rightarrow \{0, \ldots, N_e-1\} \) for edges.

Each node \( v \in V \) is assigned a type label using \( M_v \), and its feature vector is determined by a function \( F: V \rightarrow \mathbb{N}^{2N_e} \). The feature vector \( F(v) \) for a node \( v \) comprises the count of incoming and outgoing edges of each type. Specifically, for \( i \in \{0, \ldots, N_e-1\} \), the feature vector includes:
\begin{equation}
    a_i = |\{e \in E_{\text{in}}(v) : M_e(\text{type}(e)) = i\}|
    \label{eq:incoming_edges}
\end{equation}
\begin{equation}
    a_{i+N_e} = |\{e \in E_{\text{out}}(v) : M_e(\text{type}(e)) = i\}|
    \label{eq:outgoing_edges}
\end{equation}

where \( a_i \) represents the count of incoming edges of type \( i \), and \( a_{i+N_e} \) denotes the count of outgoing edges of the same type. The final feature vector \( F(v) \) is represented in Equation \ref{eq:feature_vector}:

\begin{equation}
    F(v) = [a_0, a_1, \ldots, a_{N_e-1}, a_{N_e}, a_{N_e+1}, \ldots, a_{2N_e-1}]
    \label{eq:feature_vector}
\end{equation}

\subsubsection{GNN training:}
Our GCN model is engineered to predict complex node relationships within a graph, using node types and provenance graph structures as inputs. Its primary goal is to accurately predict feature vectors for each node by leveraging their connections and positions within the graph. The architecture features three graph convolutional layers, which aggregate and transform node features according to their connectivity. Each convolution operation is defined as:

\begin{equation}
H^{(l+1)} = \sigma(\tilde{D}^{-\frac{1}{2}} \tilde{A} \tilde{D}^{-\frac{1}{2}} H^{(l)} W^{(l)})
\label{eq:convolution}
\end{equation}

In this equation, $H^{(l)}$ represents the node feature matrix at layer $l$, $\tilde{A}$ is the adjacency matrix of the graph, $\tilde{D}$ is the degree matrix of $\tilde{A}$, $W^{(l)}$ is the weight matrix for layer $l$, and $\sigma$ denotes the activation function, ReLU (Rectified Linear Unit). This convolutional operation effectively captures the local graph topology and feature information. Following the first two convolutional layers, batch normalization and ReLU activation functions are applied, with dropout used for regularization purposes. The model's predicted feature vector can be represented in Equation \ref{eq:finallayer}:

\begin{equation}
    \hat{F}(v) = H^{(final)}W^{(final)}
    \label{eq:finallayer}
\end{equation}

Then, the Mean Squared Error (MSE) is used to assess the difference between the model's prediction and the actual feature vector $F(v)$ for each node in the graph. Formally, the MSE loss is defined in Equation \ref{eq:MSE}:

\begin{equation}
\text{MSE} = \frac{1}{n} \sum_{v \in V} | F(v) - \hat{F}(v) |^2
\label{eq:MSE}
\end{equation}

Here, $n$ represents the total number of nodes in the graph G, $F(v)$ denotes the actual feature vector of node v, and $\hat{F}(v)$ is the feature vector predicted by the model for node $v$.

\subsubsection{Abnormal node identification:}
In the final phase of our model, we focus on identifying abnormal nodes by analyzing deviations from expected behavior. The model determines whether a node is anomalous by calculating the error value based on the absolute difference between the predicted feature vector ${F}_p(v)$ and the actual feature vector $F(v)$. The error score for node $v$ is computed as follows:

\begin{equation}
\text{error}(v) = | \hat{{F}}(v) - F(v) |
\label{eq:error}
\end{equation}

By setting an appropriate threshold, we distinguish between normal node behavior and significant deviations indicative of anomalies. This threshold is determined using a data-driven approach, where we analyze the distribution of error values in the validation set to select a value that optimally balances accuracy with the minimization of false positives and negatives in anomaly detection.

\subsection{Secure Model Training through FL and PHC}

Our model employs a dual-key mechanism, entailing a public key for encryption and a private key for decryption, specifically used by the PHC. A specialized entity, referred to as the Key Generation Authority (KGA), undertakes the task of generating these keys. During the setup phase, this KGA chooses two distinct large prime numbers, \(p\) and \(q\), ensuring \(p \neq q\) and the greatest common divisor (gcd) of \(pq\) and \((p-1)(q-1)\) is 1. Subsequently, the KGA defines \(N = pq\) and computes Carmichael's function as \(\lambda(N) = \text{lcm}(p-1, q-1)\). A random integer \(g\) is then chosen from \(\mathbb{Z}_{N^2}^*\) such that \(\text{gcd}(L(g^\lambda \mod n^2), N) = 1\). Here, the function \(L(x)\) is formulated as \(L(x) = \frac{x - 1}{N}\). The encryption public key is thus represented as \((N,g)\), and the decryption private key is \((\lambda, \mu)\), where \(\mu\) is a distinct integer satisfying \(\mu = L(g^\lambda \bmod N^2)^{-1} \mod N\). These keys are disseminated among organizations while being kept hidden from the server.

As each federated communication cycle commences, organizations retrieve the global aggregated updates from the preceding round from the server, and update their local model parameters according to the Equation \ref{eq:Wt+1old}:

\begin{equation}
    W_{t+1}^{old} = W_{t}^{old} + U_{t}^{global}
    \label{eq:Wt+1old}
\end{equation}

In this equation, \(W_{t+1}^{old}\) symbolizes the model updated for the round \(t+1\), with \(W_{t}^{old}\) and \(U_{t}^{global}\) representing the previous round's final organization model and aggregated updates, respectively. The organizations engage in local training on their datasets, developing the GNN model and calculating the new parameter vector \(W_{new}^{t+1}\). The updated vector for the local model, \(U_{t+1}\), intended for subsequent server transmission, is derived as \(U_{t+1} = W_{t+1}^{\text{new}} - W_{t+1}^{\text{old}}\).

Following the reception of keys from the KGA, each organization, indicated as organization \(k\), executes the PHC algorithm. This involves generating a random number \(r^k_{t+1}\) in \(Z^*_{n^2}\) where \(0 < r_{t+1}^k < n\), converting the update vector of each organization into ciphertext, expressed as \(C_{t+1}^k = g^{U_{t+1}^k} \cdot r^k_{t+1} \mod n^2\).

Post-encryption, the organization sends the ciphertext \(C_{t+1}^k\) to the server for aggregation and generation of global model updates. Utilizing the PHC, the product of two ciphertexts results in a decryption equivalent to the sum of their original messages. The server, upon collecting all organization ciphertexts, applies the Federated Averaging (FedAvg) algorithm and PHC for aggregating updates. In the final phase, the server distributes the aggregated global updates back to the organizations. Each organization then employs the private keys to execute the decryption algorithm, obtaining the final results and updating the global model for the training round \(t+2\), as shown:

\begin{equation}
U_{(t+1)}^{\text{global}} = L \left(\left( C_{(t+1)}^{\text{global}}\right)^{\lambda} \mod n^2 \right)\mu \mod n
\end{equation}

\begin{equation}
W_{t+2}^{old} = W_{t+1}^{old} + U_{t+1}^{{global}}
\end{equation}

Here, $W_{t+2}^{old}$ is the updated global model at the commencement of round $t+2$. This iterative process continues until the global model achieves convergence. 

\section{Performance Evaluation}\label{sec:Performance Evaluation}

\subsection{Experimental Setup}
We used PyTorch 2.0.1 and CUDA 11.8 with an Intel i7-13700F CPU and NVIDIA GeForce RTX 4070 GPU. Our dataset's training set (60\%) contains only normal data, while the validation and test sets (20\% each) include both normal and anomalous data, facilitating effective model evaluation and threshold tuning. The model, configured with a 0.5 learning rate and an Adam optimizer, underwent training across three rounds, each involving 100 epochs per organization. We tested scalability by varying the number of participating organizations from 2 to 10, observing stable performance with minimal fluctuations. Data encryption was managed using the python-paillier library with a 1024-bit key \cite{PythonPaillier}. The architecture of our GNN model is detailed in Table \ref{tab:GNN}.

\begin{table}[h]
\centering
\caption{Parameters of the GNN Model}
\begin{tabular}{|c|c|c|}
\hline
\textbf{Layer} & \textbf{Input Dimension} & \textbf{Output Dimension} \\ \hline
GCNConv1 & 9 & 256 \\ \hline
BatchNorm1 & - & 256 \\ \hline
GCNConv2 & 256 & 128 \\ \hline
BatchNorm2 & - & 128 \\ \hline
GCNConv3 (Output Layer) & 128 & 23 \\ \hline
\end{tabular}
\label{tab:GNN}
\end{table}

\subsection{Dataset}
The dataset for this study came from the DARPA TCE3 program, during red-team versus blue-team adversarial engagements in April 2018. These engagements, designed to simulate diverse cyberattack scenarios, involved the red team conducting attacks and the blue team defending. The setup included multiple servers such as web, SSH, email, and SMB, replicating a real-world network. The scenario realism was further enhanced by simulating extensive normal user activities, including routine system administration, web browsing, and the installation of various tools. The threat descriptions and ground truth reports from DARPA provided key insights into the red team's tactics and strategies \cite{keromytis2018transparent}. Despite the dataset's large volume of events, malicious events represent only a small fraction, making anomaly and threat detection challenging due to the significant imbalance between benign and malicious activities \cite{10.1145/3450569.3463573}.

\subsection{Evaluation Metrics}
To evaluate our model, we use several metrics based on the definitions of true positives (TP), true negatives (TN), false positives (FP), and false negatives (FN). Accuracy is the proportion of correct predictions (\(\text{Accuracy} = \frac{\text{TP} + \text{TN}}{\text{TP} + \text{TN} + \text{FP} + \text{FN}}\)), Precision is the ratio of correct attack identifications to all labeled attacks (\(\text{Precision} = \frac{\text{TP}}{\text{TP} + \text{FP}}\)), and Recall is the proportion of actual attacks correctly identified (\(\text{Recall} = \frac{\text{TP}}{\text{TP} + \text{FN}}\)). The F1-score provides a balance between precision and recall (\(\text{F1-score} = 2 \cdot \frac{\text{Recall} \cdot \text{Precision}}{\text{Recall} + \text{Precision}}\)), and False Positive Rate measures the rate of benign instances incorrectly marked as attacks (\(\text{FPR} = \frac{\text{FP}}{\text{FP} + \text{TN}}\)).

\subsection{Detection Evaluation}
The results of our proposed model and the performance comparison with other anomaly detection models are presented in Table \ref{tab:model_comparison}. Our approach stands out in its unsupervised learning methodology, which enables the detection of APT attacks without prior exposure to malicious data. The model exhibits a high accuracy of 0.93 and maintains a low false positive rate of 0.06. This performance is particularly noteworthy as it is achieved in the absence of supervised learning techniques commonly employed in existing models. Due to the lack of existing research on unsupervised learning methods applied to this dataset, this study undertakes a comparative analysis with supervised learning techniques. Our model's performance was benchmarked against several state-of-the-art models, including DeepTaskAPT \cite{9724394}, Deeplog \cite{du2017deeplog}, LogShield, and BERT \cite{afnan2023logshield}, and conventional baseline models (such as Random Forest and Linear Regression) \cite{9724394}. While some of these methods may demonstrate superior performance metrics in certain aspects, they are predominantly reliant on supervised learning models. This reliance is a significant limitation, especially in addressing zero-day challenges where the model has no prior knowledge of attack patterns. Moreover, these existing models overlook the critical aspect of privacy in log data. In contrast, our model integrates FL and homomorphic encryption, addressing privacy concerns more effectively.
\begin{table}[h]
\centering
\caption{Comparison of Model Performance}
\begin{tabular}{|c|c|c|c|c|c|c|}
\hline
\textbf{Model}           & \textbf{Method}    & \textbf{Acc} & \textbf{Prec} & \textbf{Rec} & \textbf{F1-score} & \textbf{FPR} \\ \hline
P3GNN                    & Unsupervised       & 0.93         & 0.73          & 0.91         & 0.81              & 0.06         \\ \hline
DeepTaskAPT              & Supervised         & 0.985        & N/A           & N/A          & N/A               & 0.011        \\ \hline
DeepLog                  & Supervised         & 0.83         & N/A           & N/A          & N/A               & 0.16         \\ \hline
Random Forest            & Supervised         & 0.80         & N/A           & N/A          & N/A               & 0.18         \\ \hline
Linear Regression        & Supervised         & 0.08         & N/A           & N/A          & N/A               & 0.93         \\ \hline
LogShield                & Supervised         & 0.98         & 0.99          & 0.99         & 0.984             & N/A          \\ \hline
BERT                     & Supervised         & 0.89         & 0.91          & 0.95         & 0.90              & N/A          \\ \hline
\end{tabular}
\label{tab:model_comparison}
\end{table}

\subsection{Computational Complexity Analysis of P3GNN}

The complexity of the P3GNN model arises from data extraction, graph construction, GNN training, and anomaly detection. Incorporating FL and PHC adds to this complexity by introducing encryption and decryption overheads, crucial for ensuring data privacy in distributed networks. The base complexity of the P3GNN model, without FL and PHC, depends on the log data volume, the graph's size and structure (vertices $V$ and edges $E$), and the depth of computational requirements for GNN training. These factors contribute to a largely linear complexity, scaling with the size of the data and the graph. However, the model's efficiency in this configuration benefits from direct, unencrypted operations on data and parameters. In contrast, when FL and PHC are integrated into the model, each operation involving model parameters—especially during the GNN training phase—incurs additional complexity due to encryption and decryption. This complexity is polynomial, specifically $O(n_{\text{params}} \cdot k^3)$, where $n_{\text{params}}$ is the number of model parameters, and $k$ represents the encryption key size. This added complexity is a direct consequence of the cryptographic processes required to secure the model's distributed learning environment. 
\begin{table}[h]
\centering
\scriptsize
\caption{Comparison of Computational Complexity}
\begin{tabular}{|c|c|c|}
\hline
\textbf{Component} & \textbf{Without FL and PHC} & \textbf{With FL and PHC} \\
\hline
Data Extraction & $O(n_{\text{logs}} \cdot c)$ & $O(n_{\text{logs}} \cdot c)$ \\
\hline
Contextual Data Rep. & $O(n_{\text{vertices}} + n_{\text{edges}})$ & $O(n_{\text{vertices}} + n_{\text{edges}}) + O(n_{\text{params}} \cdot k^3)$ \\
\hline
GNN Training & $O(|V| + |E|)$ per layer & $O(|V| + |E|)$ per layer + $O(n_{\text{params}} \cdot k^3)$ \\
\hline
Abnormal Node Ident. & $O(n_{\text{vertices}})$ & $O(n_{\text{vertices}}) + O(n_{\text{params}} \cdot k^3)$ \\
\hline
\end{tabular}
\label{tab:complexity_comparison}
\end{table}
Table \ref{tab:complexity_comparison} summarizes the computational complexity of P3GNN with and without privacy preservation techniques. The added computational complexity from integrating FL and PHC into the P3GNN model is a necessary trade-off for ensuring robust privacy and security in distributed ML environments. This complexity enables collaborative training across multiple entities without exposing sensitive data, aligning with stringent data protection regulations and safeguarding against unauthorized access, making it indispensable for securely handling sensitive or confidential information.

\subsection{Interpretability of P3GNN}
This section delves into the interpretability of our model, particularly focusing on its capability to detect attacks at the node level and trace the entirety of an anomalous neighborhood involved in an attack. By leveraging GNNs, P3GNN goes beyond mere identification, enabling the dissection of attack patterns through node-level analysis.

\subsubsection{Visualization and Node-Level Analysis:}
\begin{figure}[ht]
\centering
\subfloat[Pathways and sequences of activities]{
    \includegraphics[width=0.45\textwidth]{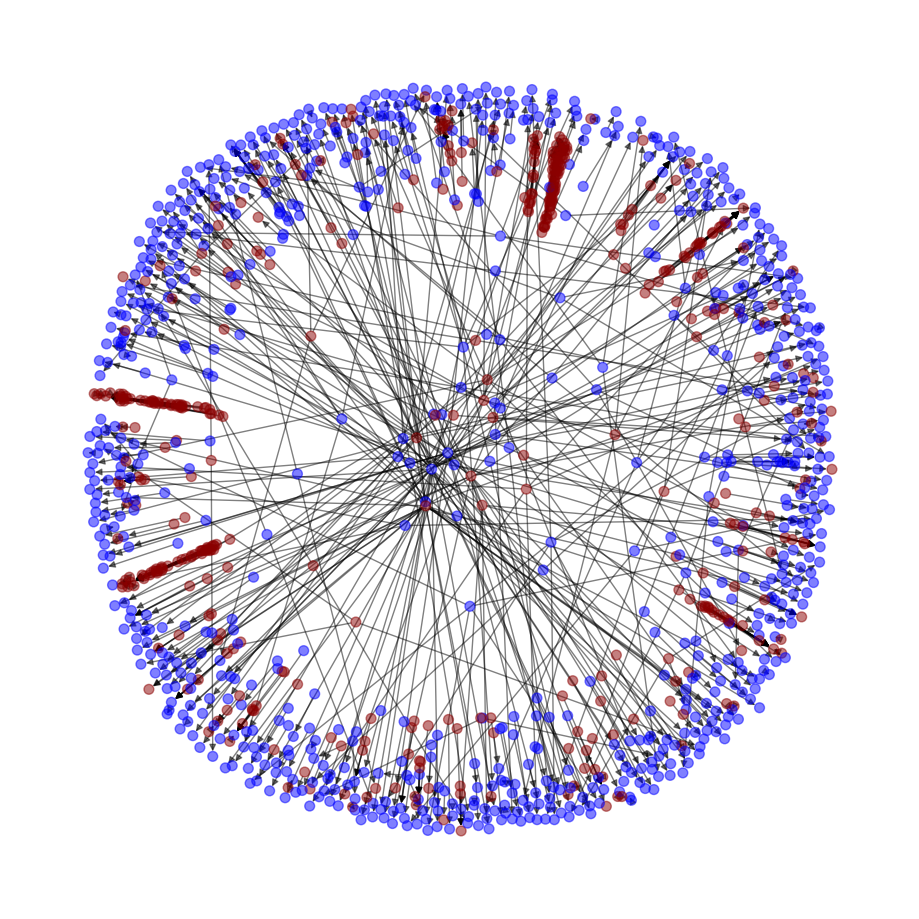}
    \label{fig3a}
}
\subfloat[Clustering of anomalous nodes]{
    \includegraphics[width=0.45\textwidth]{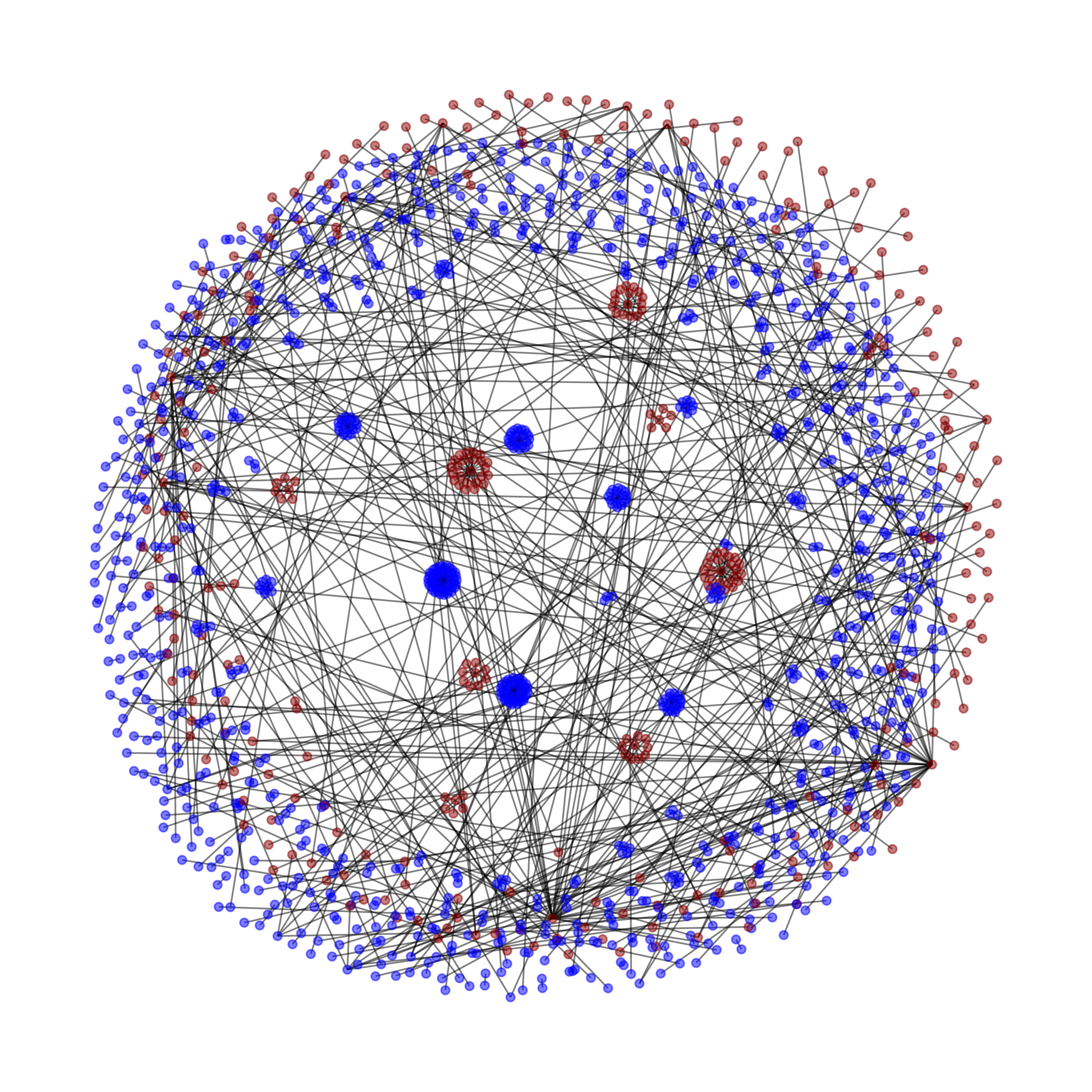}
    \label{fig3b}
}
\caption{Detailed view of anomalous node interconnectivity}
\label{fig:benignvsmalicious}
\end{figure}

P3GNN stands out for its ability to precisely identify anomalous nodes within a network. Figure \ref{fig3a} emphasizes the paths and sequences formed by anomalous nodes. The flow of malicious activities (red) amidst the benign (blue) nodes reveals the model's detection of attack trajectories, which can be critical for understanding the sequential progression and strategies of the attackers. This visualization through the capabilities of GNNs enhances interpretability, and is particularly useful for path analysis, as it enables security analysts to track potential attack trajectories and comprehend how malicious activities may propagate. It is an essential tool for formulating effective defensive strategies.

Additionally, P3GNN highlights clusters of malicious nodes that can indicate the initial breach points or critical attack vectors. Figure \ref{fig3b} illustrates a subgraph extracted from the overall network, highlighting the prominent clusters of malicious and benign nodes. In our network visualizations, clusters of nodes emerge organically due to the inherent structure and connectivity patterns within the network data. These clusters, emerging as a result of the interconnected nature of network activities, are marked by dense interconnectivity, which is a hallmark of coordinated or related malicious activities. Such clusters demand urgent investigative action. In this way, security analysts are empowered to conduct focused investigations and implement targeted security measures.

\subsubsection{Case Study: Nginx Backdoor w/ Drakon In-Memory:}

To further illustrate the model's interpretability, we analyze a specific attack incident that occurred within the DARPA network infrastructure. In this incident, the attacker targeted the system (Cadet's data), initially exploiting a vulnerability in the Nginx server using a malformed HTTP request. This led to the execution of a Drakon implant in the Nginx server’s memory, establishing a shell connection to the attacker's console through HTTP. The attacker's subsequent actions involved multiple attempts to download and elevate privileges for a micro APT implant. Despite repeated failures in privilege elevation, the attacker managed to execute the Drakon implant with root privileges and later conducted a port scan on multiple network targets to assess vulnerabilities, leaving the console connection open throughout. This attack sequence highlights several critical steps:
\begin{itemize}
    \item Initial exploitation of Nginx and Drakon implant execution.
    \item Repeated attempts to download and elevate the micro APT implant.
    \item Successful execution of the Drakon implant with elevated privileges.
    \item Network reconnaissance through port scanning.
\end{itemize}

\begin{figure}
    \centering
    \includegraphics[scale=0.08]{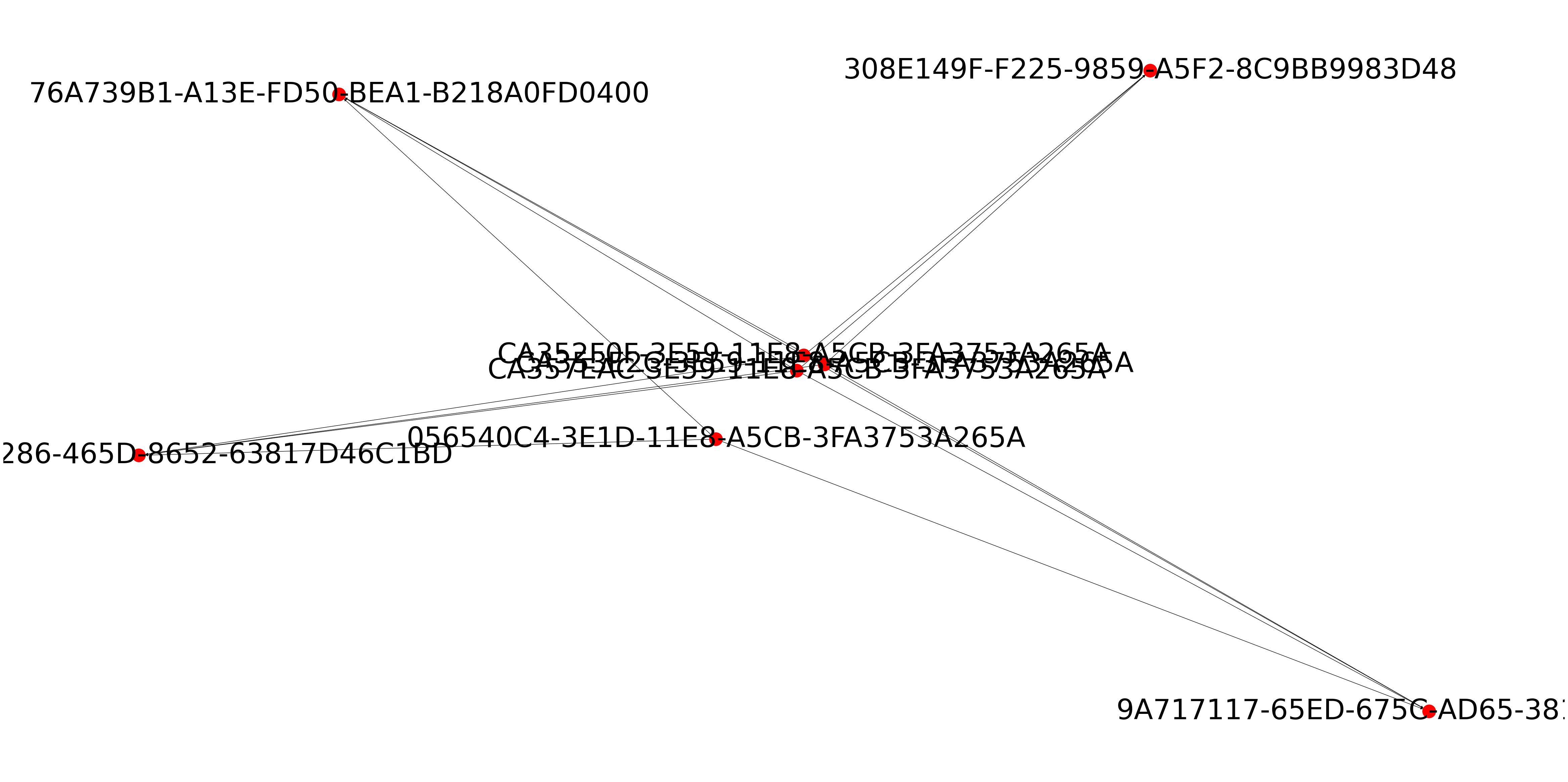}
    \caption{Visualization of attack sequence}
    \label{fig:cadets}
\end{figure}

The interpretability of our model is rooted in its ability to not just identify the occurrence of an attack but to unravel the sequence of events constituting it. When an anomalous activity is identified, the model raises alarms for these specific nodes, signaling potential security breaches. To comprehensively map out the attack's progression, we employ a depth-first search algorithm on the neighborhood of the identified anomalous nodes. In the DARPA attack case, the model successfully pinpointed each critical step, providing a comprehensive view of the attack’s progression. Figure \ref{fig:cadets} presents a graphical representation of the attack steps as identified by our model.

Our model's detection ability is further demonstrated by correlating discrete log entries with corresponding graph interactions. By examining the flow of UUIDs within the network and mapping them to recorded events, as shown in table \ref{tab:correlation}, our system constructs a narrative that aligns with known attack patterns. For instance, during the initial compromise phase, our model detected the event `EVENT\_READ` and `EVENT\_EXECUTE` in conjunction with specific UUID interactions, implying unauthorized access and potential credential compromise. As the attack transitioned to establish a foothold, modifications in process and principal changes were captured, highlighting the attacker's efforts to gain a stronger grip within the system.

Subsequently, an escalation of privileges was inferred from principal changes logged, suggesting that the attacker was attempting to gain higher-level access. This was further corroborated by the event `EVENT\_CHANGE\_PRINCIPAL` linked to UUIDs accessing sensitive system files. Our model's depth-first search illuminated these steps in the context of the larger attack framework, showcasing the depth of our analysis. In the final phases, internal reconnaissance and efforts to maintain presence were identified through repeated `EVENT\_READ` and `EVENT\_EXECUTE` actions. These steps were pivotal in the attacker's strategy to gather intelligence and ensure long-term access to the compromised system. By integrating log analysis with graph theory, our model transcends traditional detection mechanisms, offering a nuanced and dynamic approach to cybersecurity. This case study serves as a testament to the model’s efficacy in recognizing complex attack patterns and the associated network behaviors. 

\begin{table}
\centering
\caption{Correlation of Attack Phases and Log Entries with Graph Interactions}\label{tab:correlation}
\scriptsize
\begin{tabular}{|p{1.9cm}|p{3.98cm}|p{5.98cm}|} 
\hline
\makecell{Attack Phase} & \makecell{Log Entry Details} & \makecell{Graph Interaction (UUIDs)} \\
\hline
\makecell{Initial \\ Compromise} & 
\makecell{EVENT\_READ, \\EVENT\_EXECUTE \\ on /etc/login.conf, /etc/group} & 
\makecell{CA352F0F-3E59-11E8-A5CB-3FA3753A265A \\→ 9A717117-65ED-675C-AD65-38102C67C832,\\ CA357EAC-3E59-11E8-A5CB-3FA3753A265A \\→ 308E149F-F225-9859-A5F2-8C9BB9983D48}\\
\hline
\makecell{Establish \\ Foothold} & 
\makecell{EVENT\_MODIFY\_PROCESS,\\ EVENT\_CHANGE\_PRINCIPAL} & 
\makecell{CA353E2C-3E59-11E8-A5CB-3FA3753A265A \\→ 24782D99-5286-465D-8652-63817D46C1BD,\\ CA353E2C-3E59-11E8-A5CB-3FA3753A265A \\→ 76A739B1-A13E-FD50-BEA1-B218A0FD0400} \\
\hline
\makecell{Escalate \\ Privileges} & 
\makecell{EVENT\_CHANGE\_PRINCIPAL \\ indicating privilege escalation} & 
\makecell{CA353E2C-3E59-11E8-A5CB-3FA3753A265A \\→ 308E149F-F225-9859-A5F2-8C9BB9983D48} \\
\hline
\makecell{Internal \\ Reconnaissance} & 
\makecell{EVENT\_READ \\ on various system files} & 
\makecell{CA357EAC-3E59-11E8-A5CB-3FA3753A265A \\→ 9A717117-65ED-675C-AD65-38102C67C832} \\
\hline
\makecell{Maintain \\ Presence} & 
\makecell{Repeated EVENT\_READ,\\ EVENT\_EXECUTE} & 
\makecell{CA352F0F-3E59-11E8-A5CB-3FA3753A265A \\→ Various UUIDs} \\
\hline
\end{tabular}
\end{table}

\section{Conclusion}\label{sec:conclusion}
In conclusion, this paper introduced the P3GNN, a provenance graph-based model for APT detection in SDNs that integrates GCN with FL and homomorphic encryption for privacy preservation. The model effectively detects APTs without prior attack knowledge, maintaining data privacy and integrity. Our tests using the DARPA TC dataset show robust performance, achieving an accuracy of 0.93 and a false positive rate of 0.06. This indicates a significant improvement over traditional supervised learning methods, which often struggle with zero-day attacks and privacy concerns. Furthermore, the model's capability to pinpoint anomalous nodes and trace the entire attack trajectory offers invaluable insights for security analysis, addressing a critical need in AI for interpretability. In future work, we plan to broaden our study by exploring our approach’s applicability across different domains and scenarios, with a focus on empirical validation in these contexts. A key extension will involve adapting our method to defend against APT attacks across multiple hosts by shifting from host-level to network-level monitoring, enabling the detection of lateral movements in APT scenarios. Additionally, we plan to incorporate real-time data processing into the P3GNN model, enhancing its effectiveness in dynamic network environments and enabling a faster response to emerging threats. Additionally, we plan to automate the generation of Cyber Threat Intelligence reports using our P3GNN model to streamline the documentation of attack details and mitigation strategies. This automation will utilize provenance data to improve the efficiency and consistency of threat reporting and analysis in cybersecurity.


\bibliographystyle{splncs04}
\bibliography{bib}

\end{document}